# Temperature of nonextensive system:

# Tsallis entropy as Clausius entropy

Sumiyoshi Abe

*Institute of Physics, University of Tsukuba, Ibaraki 305-8571, Japan*

The problem of temperature in nonextensive statistical mechanics is studied. Considering the first law of thermodynamics and a "quasi-reversible process", it is shown that the Tsallis entropy becomes the Clausius entropy if the inverse of the Lagrange multiplier, $\beta$, associated with the constraint on the internal energy is regarded as the temperature. This temperature is different from the previously proposed "physical temperature" defined through the assumption of divisibility of the total system into independent subsystems. A general discussion is also made about the role of Boltzmann's constant in generalized statistical mechanics based on an entropy, which, under the assumption of independence, is nonadditive.

PACS number(s): 05.30.-d, 05.70.-a, 05.90.+m

Regarding nonextensive systems such as long-range interacting ones, it is essential to notice that the thermodynamic limit and the long-time limit do not commute, in general [1]. Of particular interest is the order, for which the long-time limit is taken after the thermodynamic limit, since in this case nonequilibrium stationary states, which survive for a long period of time, may be observed for a wide class of initial conditions [2]. Nonextensive statistical mechanics [1] is expected to be relevant to statistical description of such intriguing states. Therefore, it is basically unconcerned with equilibrium. Then, it is important to clarify to what extent the ordinary macroscopic thermodynamic principles can apply to such long-lived stationary states. A problem immediately arising here is the meaning of temperature of a nonextensive system, since the concept of temperature is defined through the zeroth law of thermodynamics for equilibrium. This problem has repeatedly been discussed in the literature [3-8]. In particular, the so-called "physical temperature" proposed in [3] is defined with respect to Tsallis' nonadditive entropy under the assumptions that the total system can be divided into *independent* subsystems and the total internal energy is simply given by the *sum* of those the subsystems. Later, the assumption on the properties of the internal energy was somewhat relaxed [9], but it is still not clear if such an internal-energy structure can be understood in terms of generic Hamiltonian dynamics. Generally speaking, it may not be possible to divide a nonextensive system into independent subsystems. This brings a fundamental difficulty to defining temperature in conformity with the zeroth law of thermodynamics [10] (see also [11,12]). However, as mentioned above, we are concerned not with strict equilibrium states but with nonequilibrium stationary states. There is, therefore, a possibility that it is not pertinent to presume that temperature in nonextensive statistical mechanics should also be defined through the standard zeroth law, which is associated with the long-time limit before the thermodynamic limit.

In this paper, we study the problem of temperature in nonextensive statistical mechanics from the viewpoint of the fundamental thermodynamic principles except the zeroth law. We consider a "quasi-reversible process" between nonequilibrium stationary states in order to establish the relation between the quantity of heat and the Clausius entropy. It is a conceptual counterpart (in the nonequilibrium stationary situation) of a familiar quasi-static process. This scheme, in turn, enables us to identify what temperature is. It is shown that the Tsallis entropy [13] becomes identical to the Clausius entropy if the inverse of the Lagrange multiplier, $\beta$, associated with the constraint on the internal energy, is regarded as the system temperature. Therefore, it is different from "physical temperature" in [3] defined through the zeroth law as well as the assumption of divisibility of the total system into independent subsystems. We also develop a general discussion about the role of Boltzmann's constant for Tsallis' nonadditive entropy, and make a comment on a recent work [14], which shows that introduction of specific correlation between subsystems can make the total Tsallis entropy be the sum of the sub-entropies.

As mentioned above, the bases put on our discussion are the fundamental thermodynamic principles. Specifically, only the first law and Clausius' definition of the thermodynamic entropy are employed, and no other details such as the Legendre transformation structure or less fundamental thermodynamic relations are imposed.

We first recall the first law in nonextensive statistical mechanics formulated in [15] (in which the second law of thermodynamics is also established). Given the system Hamiltonian, $H$, with its value $\varepsilon_i$ in the $i$th state of the system, the internal energy in nonextensive statistical mechanics is given by the $q$-expectation value [16,17]

$$U_q = \langle H \rangle_q = \frac{\sum_i \varepsilon_i (p_i)^q}{\sum_j (p_j)^q}. \tag{1}$$

Here, $p_i$ stands for the probability distribution describing a nonequilibrium stationary state (i.e., the maximum Tsallis entropy state) of the nonextensive system under consideration, and is given by the so-called $q$-exponential distribution

$$p_i = \frac{1}{Z_q}\left[1 - (1-q)(\beta/c_q)(\varepsilon_i - U_q)\right]_+^{1/(1-q)}, \tag{2}$$

$$Z_q = \sum_i \left[1 - (1-q)(\beta/c_q)(\varepsilon_i - U_q)\right]_+^{1/(1-q)}, \tag{3}$$

$$c_q = \sum_i (p_i)^q = (Z_q)^{1-q}, \tag{4}$$

with the notation, $[a]_+ = \max\{0, a\}$. In these equations, $\beta$ is the Lagrange multiplier associated with the constraint on the internal energy in maximum Tsallis entropy principle. Taking the variation of Eq. (1), we have

$$\delta U_q = \langle \delta H \rangle_q + q \frac{\sum_i \varepsilon_i (p_i)^{q-1} \delta p_i}{c_q} - q \langle H \rangle_q \frac{\sum_i (p_i)^{q-1} \delta p_i}{c_q}. \tag{5}$$

This variation should be regarded as the description of changes of the physical quantities through a quasi-reversible process connecting two nonequilibrium stationary states that are infinitesimally distinct each other. Identifying the first term on the right-hand side with the work done by the external environment to the system, that is,

$$\delta' W_q = -\langle \delta H \rangle_q, \tag{6}$$

we obtain the following expression for the quantity of heat:

$$\delta' Q_q = q \frac{\sum_i (\varepsilon_i - U_q)(p_i)^{q-1} \delta p_i}{c_q}$$

$$= \frac{q}{1-q} \frac{c_q}{\beta} \sum_i \left[1-(1-q)(\beta/c_q)(\varepsilon_i - U_q)\right]^{-1} \delta p_i, \tag{7}$$

where Eq. (2) has been used in the last equality. Consequently, we establish the first law in nonextensive statistical mechanics

$$\delta' Q_q = \delta U_q + \delta' W_q. \tag{8}$$

Now, the crucial question is if $\delta' Q_q$ can be related to the Clausius entropy. In other words, what temperature can mediate between the quantity of heat and the entropy? To answer to this question, let us examine the Tsallis entropy indexed by the positive value of $q$

$$S_q[p] = \frac{1}{1-q}\left[\sum_i (p_i)^q - 1\right]. \tag{9}$$

Here, Boltzmann's constant, $k_B$, is set equal to unity, but later we will introduce it, explicitly. Taking its variation and using Eq. (2), we have

$$\delta S_q = \frac{q}{1-q} c_q \sum_i \left[1-(1-q)(\beta/c_q)(\varepsilon_i - U_q)\right]^{-1} \delta p_i. \tag{10}$$

Comparing Eqs. (7) and (10), we arrive at the result

$$\delta S_q = \beta\, \delta' Q_q. \tag{11}$$

This implies that *the Tsallis entropy becomes the Clausius entropy if $\beta^{-1}$ is identified with temperature*, which is notably different from the physical temperature discussed in [3]. It should also be noticed that such identification may be in conformity with Kelvin's definition of thermodynamic absolute temperature. Eq. (11) sets a relation between two concepts: temperature and entropy. Therefore, one of the two is left unspecified.

Existence of the relation in Eq. (11) as ascertained by Carathéodory's theorem in ordinary thermodynamics is crucially based on the concept of additivity [18] (additivities of the internal energy and the Clausius entropy are also indispensable for formulating the zeroth law). Here, additivity means that when the total system is divided to two *independent* subsystems, *A* and *B*, the internal energy and the entropy are the sums of those of *A* and *B*. However, such divisibility cannot be realized if there exists strong correlation, as typically in a complex/nonextensive system. It may lead to a nontrivial aspect in nonextensive statistical mechanics, which requires further clarification. Let us see this through the following rhetorical question. Assume the joint probability, $p_{ij}(A, B)$, be factorized: $p_{ij}(A, B) = p_i(A) p_j(B)$, i.e., there is no correlation. Then, what happens for the Tsallis entropy in Eq. (9)? Being multiplied by Boltzmann's constant, $k_B$, it yields

$$S_q(A, B) = S_q(A) + S_q(B) + \frac{1-q}{k_B} S_q(A) S_q(B). \tag{12}$$

This relation is not acceptable, if we wish to identify the Tsallis entropy with the Clausius entropy in macroscopic thermodynamics. It is because of the explicit appearance of $k_B$ on the right-hand side. Boltzmann's constant is a unit of *microscopicness*, and therefore should not explicitly appear in macroscopic thermodynamic relations: it may appear only through the equation of state of a specific entity composed of microscopic molecules or atoms.

Thus, we are *forced* to take into account correlation between *A* and *B*. Accordingly, the joint probability has the generic Bayesian form: $p_{ij}(A, B) = p_i(A) p_{ij}(B|A)$ $= p_{ij}(A|B) p_j(B)$, where $p_{ij}(A|B)$ and $p_{ij}(B|A)$ are the conditional probabilities. Then, the total Tsallis entropy satisfies

$$S_q(A, B) = S_q(A) + S_q(B|A) + \frac{1-q}{k_B} S_q(A) S_q(B|A)$$

$$= S_q(A|B) + S_q(B) + \frac{1-q}{k_B} S_q(A|B) S_q(B), \tag{13}$$

where $S_q(A|B)$ and $S_q(B|A)$ are the conditional Tsallis entropies defined in [19]. However, Eq. (13) still contains Boltzmann's constant, explicitly.

In a recent intriguing work [14], it has been discussed that a limited class of the pattern of correlation can make the total Tsallis entropy be

$$S_q(A, B) = S_q(A) + S_q(B), \quad (14)$$

which means that such correlation leads to $S_q(B|A) + \frac{1-q}{k_B} S_q(A) S_q(B|A) = S_q(B)$, for example. Now, Eq. (14) does *not* contain Boltzmann's constant, explicitly. This is certainly a preferable feature and is physically natural from the thermodynamic viewpoint (see [20]). In particular, it may offer a possibility of generalizing the conventional zeroth law of thermodynamics to the case of the nonadditive internal energy, or the total Hamiltonian of a generic form

$$H(A, B) = H(A) + H(B) + H_{int}(A, B). \quad (15)$$

The situation is, however, not clear yet. To further elaborate the issue, let us employ the symmetric two-level system considered in [14] for the sake of simplicity. The joint distribution proposed in [14] is given by

$$p_{11}(A, B) = r^2 + \kappa, \tag{16}$$

$$p_{12}(A, B) = r(1-r) - \kappa, \tag{17}$$

$$p_{21}(A, B) = r(1-r) - \kappa, \tag{18}$$

$$p_{22}(A, B) = (1-r)^2 + \kappa, \tag{19}$$

where $0 \leq r \leq 1$, and $\kappa = \kappa(r)$, responsible for correlation, is depicted in Fig.1 in [14] with respect to $r$ for some values of the entropic index, $q$. Clearly, the pattern of correlation in Eqs. (16)-(19) is not trivially compatible with the $q$-exponential distribution in Eq. (2) with the generic form of the Hamiltonian in Eq. (15). Even without the interaction term, there remains correlation induced by statistical nonextensivity [21]. This incompatibility gives an important indication regarding interpretation of nonextensive statistical mechanics. Nonextensive statistical mechanics is supposed to be a theory for systems whose phase-space configurations have nontrivial clustering/hierarchical structures with strong correlation and broken ergodicity. Though such emergent properties are certainly induced by dynamics, it does not seem to be appropriate to identify whole correlation with interaction, since correlation may be of the higher-order property built upon dynamics. In nonextensive statistical mechanics, there is the interplay between interaction and the entropic index, $q$. The latter is supposed to effectively describe the emergent properties.

The discussion in [14] also leads to the following natural question: should all complex/nonextensive systems obey the specific pattern of correlation as in Eqs. (16)-(19)? It is unlikely that the answer is affirmative. There may exist freedom in identifying a relevant generalized entropy. It is known in the literature that there are a number of generalized entropies such as the quantum-group entropy [22], the $\kappa$-entropy [23], the two-parameter deformed entropy [24] or even a highly general entropy [25]. In particular, the quantum-group entropy and the $\kappa$-entropy are concave, Lesche-stable [26,27], and satisfy the $H$-theorem [25]. In addition, like the Tsallis entropy [28], they satisfy the extended Pesin identity and yield the finite entropy production rate at the edge of chaos of certain nonlinear dynamical systems [29]. To be able to be expressed as the sum of sub-entropies, these generalized entropies require patterns of correlation, which are different from that of the Tsallis entropy, e.g., in Eqs. (16)-(19), and might cover classes of systems. Clearly, further investigation is needed along this line.

In conclusion, we have revisited the problem of temperature in nonextensive statistical mechanics by considering the first law of thermodynamics and quasi-reversible processes. We have seen that the Tsallis entropy becomes the Clausius entropy if the inverse of the Lagrange multiplier, $\beta$, is identified with the temperature, which is notably different from the previously proposed "physical temperature" defined through the assumption of divisibility of the total system into independent subsystems. We have also developed a general discussion about the role of tBoltzmann's constant in generalized statistical mechanics based on nonadditive entropy.

**Acknowledgment**

This work was supported in part by the Grant-in-Aid for Scientific Research of Japan Society for the Promotion of Science.